\documentclass[runningheads]{llncs}
\usepackage{amsmath}
\usepackage{dsfont}
\usepackage{graphicx}
\usepackage{tikz}
\usepackage{multirow}
\usepackage{array}
\usepackage{tabularx,booktabs}
\usepackage{subfigure}
\usepackage{float}
\usetikzlibrary{arrows.meta}
\usepackage{color}
\usepackage{hyperref}

\newcolumntype{Y}{>{\centering\arraybackslash}X}
\begin{document}
\title{Impact of Adversarial Examples on \\ Deep Learning Models for\\Biomedical Image Segmentation}

\titlerunning{Adversariality and Deep Learning Models for Biomedical Segmentation}

\author{Utku Ozbulak\inst{1,3}, Arnout Van Messem\inst{2,3}, Wesley De Neve\inst{1,3}}
\authorrunning{U. Ozbulak et al.}
\institute{Department of Electronics and Information Systems, Ghent University, Belgium\and Department of Applied Mathematics, Computer Science and Statistics,\\ Ghent University, Belgium\and
Center for Biotech Data Science,Ghent University Global Campus, South Korea
\email{\{firstname.lastname\}@ugent.be}}

\maketitle
\begin{abstract}
Deep learning models, which are increasingly being used in the field of medical image analysis, come with a major security risk, namely, their vulnerability to adversarial examples. Adversarial examples are carefully crafted samples that force machine learning models to make mistakes during testing time. These malicious samples have been shown to be highly effective in misguiding classification tasks. However, research on the influence of adversarial examples on segmentation is significantly lacking. Given that a large portion of medical imaging problems are effectively segmentation problems, we analyze the impact of adversarial examples on deep learning-based image segmentation models. Specifically, we expose the vulnerability of these models to adversarial examples by proposing the Adaptive Segmentation Mask Attack (ASMA). This novel algorithm makes it possible to craft targeted adversarial examples that come with (1) high intersection-over-union rates between the target adversarial mask and the prediction and (2) with perturbation that is, for the most part, invisible to the bare eye. We lay out experimental and visual evidence by showing results obtained for the ISIC skin lesion segmentation challenge and the problem of glaucoma optic disc segmentation. An implementation of this algorithm and additional examples can be found at \url{https://github.com/utkuozbulak/adaptive-segmentation-mask-attack}.
\end{abstract}

\section{Introduction}
\let\thefootnote\relax\footnotetext{Accepted for the 22nd International Conference on Medical Image Computing and Computer Assisted Intervention (MICCAI-19).}
Recent studies adopt deep learning models at a quick pace to solve image-related problems for medical data sets. Provided that (1) labor expenses (i.e., salaries of nurses, doctors, and other relevant personnel) are a key driver of high costs in the medical field and that (2) increasingly super-human results are obtained by machine learning systems, an ongoing discussion is to replace or augment manual labor with automation for a number of medical diagnosis tasks~\cite{medical_classification_adv}. However, a recent development called \textit{adversarial examples} showed that deep learning models are vulnerable to gradient-based attacks~\cite{LBFGS}. These so-called adversarial examples are now considered a major security flaw, since they allow for the use of possible fraud schemes (e.g., for insurance claims) when deep learning models are deployed for clinical tasks~\cite{medical_classification_adv}.

\begin{figure}[t]
\centering
\begin{tikzpicture}
\centering
\node[inner sep=0pt] (russell) at (0.5, 0)  
    {\includegraphics[width=.17\textwidth]{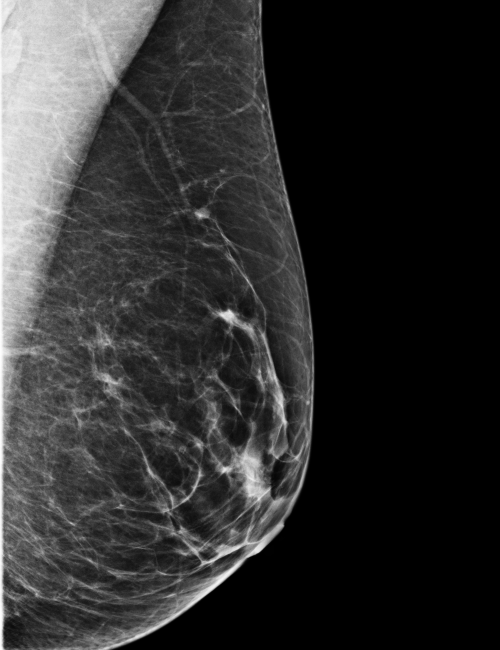}};
\node[inner sep=0pt] (whitehead) at (4.5, 0)  
    {\includegraphics[width=.17\textwidth]{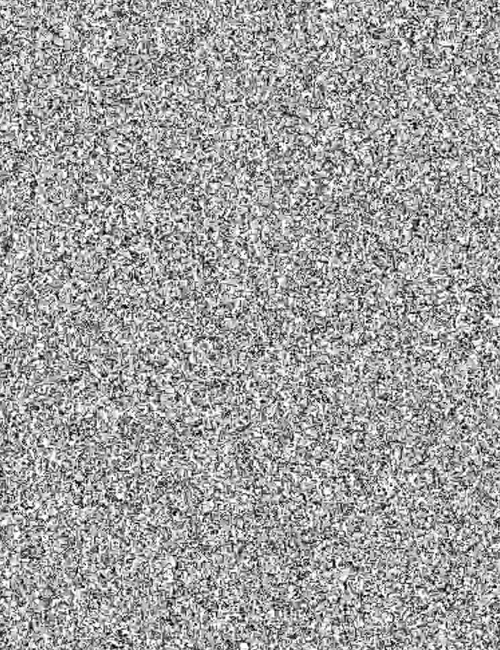}};
\node[inner sep=0pt] (whitehead) at (8.2, 0)  
    {\includegraphics[width=.17\textwidth]{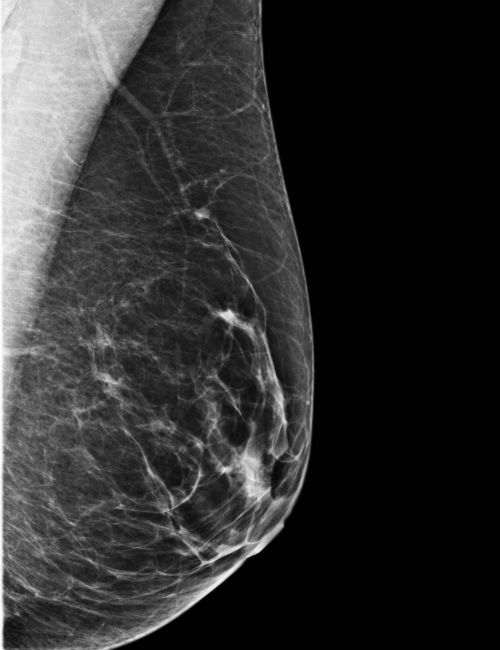}};
    \node[align=left] at (2.5, 0) {$+ \, \, \, 0.01\times$};
    \node[align=left] at (6.3, 0) {$=$};
    \node[align=center] at (0.5, -2) {Genuine Image\\ Prediction: Cancer \\ Confidence: $0.95$};
    \node[align=center] at (4.5, -1.9) {Perturbation\\(Enhanced $\times \, 100$ )};
    \node[align=center] at (8.2, -2) {Adversarial Example\\ Prediction: Healthy \\ Confidence: $0.99$};
\end{tikzpicture}
\caption{A genuine image, initially classified as \textit{cancer} with $0.95$ confidence by a deep learning model, is perturbed to become an adversarial example. This adversarial example is then classified as \textit{healthy} with $0.99$ confidence by the same model.}
\label{fig:adv_breast}
\end{figure}
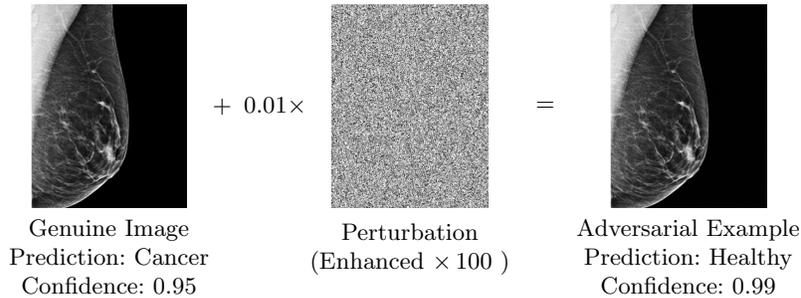

The study of adversarial examples started with \cite{LBFGS}, in which the authors observed that small pixel modifications led to large changes in the prediction. Ever since, numerous attempts were made to mitigate the impact of adversarial examples and to fix this so-called security flaw, only to be found ineffective by subsequent studies~\cite{DBLP:journals/corr/CarliniW17}. Although the effects of adversarial examples are largely studied for non-medical datasets, it was also shown that classification problems in medical imaging datasets are of no exception to this exploit~\cite{medical_classification_adv}. An adversarial example in the context of breast cancer classification is given in Figure~\ref{fig:adv_breast}.

In the field of non-medical imaging, pixel by pixel detail is most of the time not task critical. As a result, segmentation problems are often expressed as detection or localization problems~\cite{pascal_voc}. However, in medical imaging, precision is of utmost importance. Therefore, instead of detection or localization, segmentation covers a large portion of medical imaging problems~\cite{medical_seg1}.

Even though adversarial examples are studied extensively in the context of classification problems, it is only recently that studies started to investigate this phenomenon in the context of segmentation problems~\cite{arnab2018robustness,xie2017adversarial}. Thus far, in terms of segmentation, adversarial examples have been studied for the Pascal VOC~\cite{pascal_voc} and Cityscapes~\cite{Cordts2016Cityscapes} data sets, with a sole adversarial example generation method proposed in \cite{xie2017adversarial}. In particular, the Dense Adversary Generation (DAG) algorithm proposed in \cite{xie2017adversarial} aims to force deep learning models to segment all pixels wrong. Although the authors report that their algorithm is able to create adversarial examples in the context of segmentation, the resulting segmentation predictions, especially in the medical domain, are not realistic (i.e., the shape of the prediction immediately gives away that the input has been tampered with, since the prediction shape is not specified).

In this study, we focus on analyzing adversarial examples in the context of medical image segmentation problems. We demonstrate that adversarial examples indeed exist when dealing with medical image segmentation problems, discussing examples that have been obtained for glaucoma optic disc segmentation~\cite{Glucoma_seg} and ISIC skin lesion segmentation~\cite{ISIC_dataset}. Furthermore, we introduce a novel algorithm that is tailored to produce targeted adversarial examples for image segmentation problems.

To the best of our knowledge, this is the first study that analyzes the impact of adversarial examples, not only in the context of medical imaging but also in the case where the nature of the prediction is binary. Additionally, our algorithm is the first approach towards producing targeted adversarial examples for image segmentation that leads to a convincing prediction shape of choice, thus exposing a large security threat for image segmentation models. Our algorithm, while achieving targeted predictions with a high success rate, modifies the original image so subtly that the modifications on the original image are, for the most part, invisible to the bare eye.

\section{Notation and Framework}
\label{Notation}
In this section, we explain the datasets, the deep learning models, the notation, and the evaluation metrics used throughout the paper.

\textbf{Framework} \textemdash \,
In order to show the effectiveness of the proposed approach, we evaluate our attack on two datasets: the first one is the glaucoma optic disc segmentation dataset \cite{Glucoma_seg} and the second one is the ISIC skin lesion segmentation dataset \cite{ISIC_dataset}. The results reported in Section~\ref{Experiments_Section} are obtained for two separate U-Net models which is one of the most used architectures in the field of medical segmentation~\cite{DBLP:journals/corr/RonnebergerFB15}. These U-Net models have been trained on the two aforementioned datasets, achieving a segmentation effectiveness comparable to the state-of-the-art.

\textbf{Neural Network Notation} \textemdash \,
We define the forward pass in a neural network as a function $g$ with the weights and parameters of the same network detailed as $\theta$. This function takes an input image $\mathbf{X}$ of size $C\times H \times W$, with $C$, $H$, and $W$ representing the number of channels (i.e., $1$ for grayscale images, $3$ for colored images), the height, and the width of the input image, respectively. In this setting, $g(\theta, \mathbf{X})$ represents the prediction of this neural network for a given input image $\mathbf{X}$. The prediction is of size $M\times H \times W$, where $M$ is the total number of classes (e.g., two in the case of binary segmentation). We define $\mathbf{Y} := \arg \max_M (g(\theta, \mathbf{X}))$ as the prediction of the neural network after discretization, containing prediction classes (i.e., values from $0$ to $M-1$) per pixel and having a size of $H \times W$.

\textbf{Distance Metrics} \textemdash \, 
When an adversarial example is generated, a distance metric is required in order to measure the difference between the original image and the generated image. Following previous studies~\cite{CW_Attack}, we use the Euclidian distance ($L_{2}$) and the max distance ($L_{\infty}$), with the latter measuring the maximum change for a single pixel among all pixels.

\textbf{Accuracy Metrics} \textemdash \, 
In order to quantify the accuracy of the adversarial example generation, we calculate the intersection over union (IoU) and the pixel accuracy (PA) between the target mask and the predicted segmentation mask associated with the adversarial example produced. In our settings where the background label is selected as $0$, IoU and PA are defined as follows:
\begin{align*}
\text{IoU}(\mathbf{Y}^1, \mathbf{Y}^2) = 
{\displaystyle\sum_{i,j} A_{i,j} \cap  B_{i,j} } \Big/
{\displaystyle\sum_{i,j} B_{i,j} } \,,\,\,\,\text{PA}(\mathbf{Y}^1, \mathbf{Y}^2) = 
{\displaystyle\sum_{i,j} A_{i,j}} \Big/ \,
(H \times W), \,\,
\end{align*}
for $(i, j) \in (\{1, \cdots, H \}, \{1, \cdots, W \})$ and where $A_{i,j} = \mathds{1}_{\{ \mathbf{Y}^{2}_{i,j} \,=\, \mathbf{Y}^{1}_{i,j}\}}$ and $B_{i,j} = \mathds{1}_{\{ \mathbf{Y}^{2}_{i,j} \, + \, \mathbf{Y}^{1}_{i,j} \neq 0\}}$, with $\mathds{1}$ representing the indicator function.

\section{Generating Adversarial Examples}
\label{Generating Adversarial Examples}
To justify our design choices, we briefly highlight the differences between classification and segmentation in terms of adversarial example generation, before detailing our algorithm for generating adversarial examples for image segmentation.

\textbf{Adversarial Target} \textemdash \, 
In classification, the adversarial target is often a single class~\cite{IFGS}. However, in segmentation, the target is a mask. Thus, the aim is to change the prediction of not just one but of a large number of labels.

\textbf{Perturbation Multiplier} \textemdash \, 
In the case of classification (when a single class is targeted), the optimization is influenced by only one source. However, in segmentation, as a consequence of the adversarial target being a mask, the optimization is influenced by a large number of sources (i.e., individual pixels). As a result, the perturbation multiplier (i.e., the learning rate) is harder to tune.

Keeping the aforementioned differences in mind and building upon the knowledge acquired from studying adversarial examples in classification problems, we propose the Adaptive Segmentation Mask Attack (ASMA), a novel algorithm for generating targeted adversarial examples for image segmentation models.

\subsection{Adaptive Segmentation Mask Attack (ASMA)}
\label{Adaptive Mask Segmentation Attack}
As a starting point, we use the standard way of adversarial example generation, which is defined as follows:
\begin{align*}
    &\text{minimize} \,\,  ||\, \mathbf{X} \, - \, (\mathbf{X} + \mathbf{P}) \,||_{2} \,, \\
    &\text{such that} \,\,  \arg \max \big(g(\theta,  (\mathbf{X} + \mathbf{P})) \big)= \mathbf{Y}^{A} \,\, , \,\,\, (\mathbf{X} + \mathbf{P}) \in [0, 1]^{ C \times H \times W} \,.
\end{align*}

\begin{figure}[t]
\centering
\begin{tikzpicture}
\def\setx1{-5}
\def\sety1{0.3}
\node[inner sep=0pt] (c1) at (\setx1, \sety1)  
    {\includegraphics[width=1.6cm]{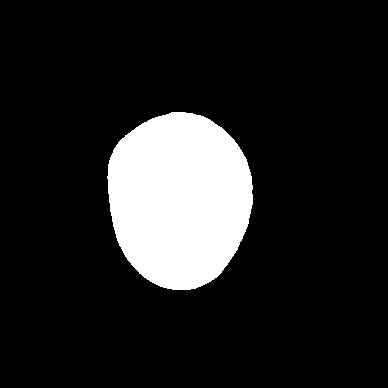}};
\node[inner sep=0pt] (t1) at (\setx1 + 2,\sety1)  
    {\includegraphics[width=1.6cm]{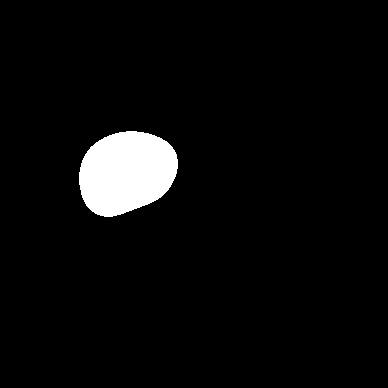}};
\node[inner sep=0pt] (t1) at (\setx1 + 4,\sety1)  
    {\includegraphics[width=1.6cm]{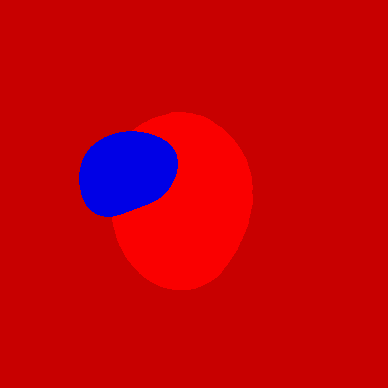}};
    
    \node[align=center] at (\setx1 , 1.7) {\tiny Initial};
    \node[align=center] at (\setx1 , 1.5) {\tiny Segmentation};
    \node[align=center] at (\setx1, 1.3) {\tiny Prediction};
    \node[align=center] at (\setx1+2, 1.7) {\tiny Target};
   \node[align=center] at (\setx1+2, 1.5) {\tiny Adversarial};
   \node[align=center] at (\setx1+2, 1.3) {\tiny Mask};
    \node[align=center] at (\setx1+4 , 1.7) {\tiny Static Adversarial};
    \node[align=center] at (\setx1+4 , 1.5) {\tiny Optimization};
    \node[align=center] at (\setx1+4, 1.3) {\tiny Mask};
    
\def\sety1{0}      
    \node[align=center] at (1.25, 1.4) {\tiny Prediction};
    \node[align=center] at (1.25, 1.2) {\tiny During};
    \node[align=center] at (1.25, 1) {\tiny Optimization};
    \node[align=center] at (1.2, 0.4) {\tiny Dynamic};
    \node[align=center] at (1.2, 0.2) {\tiny Adversarial};
    \node[align=center] at (1.2, 0) {\tiny Optimization};
    \node[align=center] at (1.2, -0.2) {\tiny Mask};
    
\def\setx1{2.6}
\node[inner sep=0pt] (c1) at (\setx1, \sety1)  
    {\includegraphics[width=1.1cm]{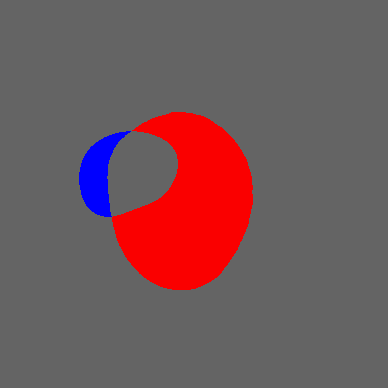}};
\node[inner sep=0pt] (t1) at (\setx1,\sety1+1.2)  
    {\includegraphics[width=1.1cm]{opt_ims/from1.png}};
    
\def\setx1{3.8}
\node[inner sep=0pt] (c1) at (\setx1, \sety1)  
    {\includegraphics[width=1.1cm]{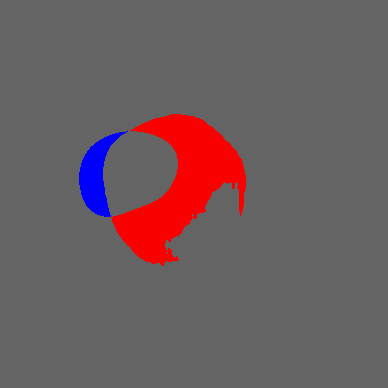}};
\node[inner sep=0pt] (t1) at (\setx1,\sety1+1.2)  
    {\includegraphics[width=1.1cm]{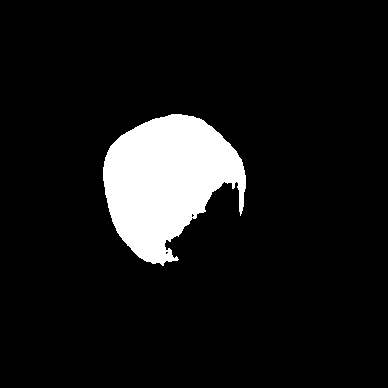}};
    
\def\setx1{5}
\node[inner sep=0pt] (c1) at (\setx1, \sety1)  
    {\includegraphics[width=1.1cm]{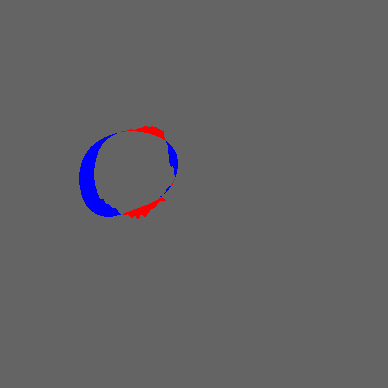}};
\node[inner sep=0pt] (t1) at (\setx1,\sety1+1.2)  
    {\includegraphics[width=1.1cm]{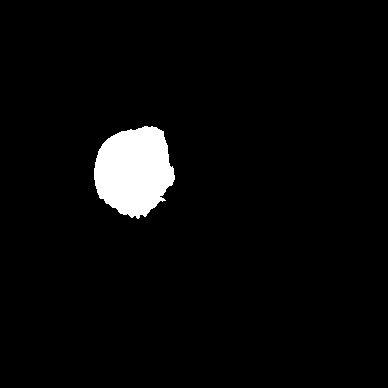}};
\end{tikzpicture}
\caption{An example adversarial optimization mask used by the static and the dynamic mask approaches, visualized in terms of the difference between the initial prediction and the targeted prediction.
}
\label{fig:static_dynamic_mask}
\end{figure}

This equation iteratively aims at finding a small perturbation $\mathbf{P}$ that is sufficient to change the prediction of the model to $\mathbf{Y}^{A}$, which we will refer to as the target adversarial mask, while keeping the $L_2$ distance between the original image $\mathbf{X}$ and its adversarial counterpart $\mathbf{X} + \mathbf{P}$ minimal. In this setting, a perturbation is calculated in an iterative manner, multiplied with a constant, and then added to the image: $\mathbf{X}_{n+1} = \mathbf{X}_n + \alpha \, \mathbf{P}_n$. We now detail how the perturbation $\mathbf{P}_n$ is calculated.

\textbf{Static Segmentation Mask (SSM)} \textemdash \, In the context of adversariality, a targeted segmentation attack must (i) increase the prediction likelihood of the selected foreground pixels in the target adversarial mask (i.e., blue areas in the static adversarial optimization mask in Figure~\ref{fig:static_dynamic_mask}), while (ii) reducing the prediction likelihood of all other pixels that are not specified in the same mask (i.e.,  red areas in the static adversarial optimization mask in Figure~\ref{fig:static_dynamic_mask}). To achieve this property, we write $\mathbf{P}_n$ as a sum of perturbations as follows:
\begin{align}
\mathbf{P}_n = \displaystyle \sum_{c=0}^{M-1} \nabla_x \big( g(\theta, \mathbf{X}_n)_c \, \odot \, \mathds{1}_{\{\mathbf{Y}^{A} \, =\, c\}} \big) \,,
\label{eq:perturbations}
\end{align}
where $\odot$ denotes the Hadamard product and $\mathbf{Y}^{A}$, again, denotes the desired prediction mask for the adversarial example that contains class labels (as shown in Figure~\ref{fig:static_dynamic_mask}). $g(\theta, \mathbf{X})_c$, on the other hand, denotes the channels of the prediction made by the neural network (i.e., in the case of binary prediction, $c=0$ for background channel and $c=1$ for foreground channel).

We will now describe how we improve this approach.

\textbf{Adaptive Segmentation Mask (ASM)} \textemdash \, When the static mask approach is used to generate adversarial examples, the gradient is sourced from the same number of target pixels in the target adversarial mask at each iteration. However, during the adversarial optimization, the prediction of certain pixels may already be correct (e.g., gray areas in the dynamic masks given in Figure~\ref{fig:static_dynamic_mask}), and may thus not require any further optimization. In order to ensure that the optimization is only sourced from pixels whose predictions are not in line with the target adversarial mask, we introduce an approach that makes use of adaptive mask targeting. To achieve this property, we write $\mathbf{P}_n$ as follows:
\begin{align}
\mathbf{P}_n = \displaystyle \sum_{c=0}^{M-1} \nabla_x \big( g(\theta, \mathbf{X}_n)_c \, \odot \, \mathds{1}_{\{\mathbf{Y}^{A} \, =\, c\}}\, \odot \, \mathds{1}_{\{\arg \max_M(g(\theta, \mathbf{X}_n)) \, \neq \, c\}} \big) \,.
\label{eq:adaptive_mask}
\end{align}

Writing $\mathbf{P}_n$ this way ensures that the gradient is only sourced from pixels whose labels are different from the target adversarial mask at each iteration.

Using the proposed adaptive segmentation mask approach, the number of pixels that source the optimization process starts high and as the prediction becomes in line with the target adversarial mask, lessens gradually.

\textbf{Dynamic Perturbation Multiplier (DPM)} \textemdash \, Since the usage of ASM progressively reduces the number of pixels the optimization sources from, setting the perturbation multiplier $\alpha$ to a fixed number either causes the optimization to halt when $\alpha$ is low or causes it to create large perturbations when $\alpha$ is high. Therefore, we employ a dynamic perturbation multiplier strategy in our adaptive mask approach, using $\alpha_n =  \beta \,\times\, \text{IoU}(\mathbf{Y}^A, \mathbf{Y}_n) \,+\, \tau$, where $\beta$ and $\tau$ are parameters used to calculate the final perturbation multiplier, also taking into account the IoU score of the prediction at the $n$th iteration. This method allows increasing the value of the multiplier dynamically as the number of pixels to be optimized decreases.

We name the adversarial example generation method that incorporates the aforementioned techniques (i.e., ASM and DPM) the \textbf{Adaptive Segmentation Mask Attack} (ASMA). Note our algorithm also works in the case where the prediction is not binary.

\begin{table}[t]
\centering
\caption{Experimental results in terms of image modification and prediction mask accuracy. We highlight the $L_2$ distances and IoU overlaps to emphasize the increase in the effectiveness of the optimization technique between the first and the last version. Note that numbers listed in this table are calculated for images whose pixel values are between $0$ and $1$.}\label{tab1}
\begin{tabularx}{12cm}{ c YYYYYYYY}
\cline{2-9}
   \multicolumn{1}{c}{} 
 & \multicolumn{4}{c}{Glaucoma Dataset}  
 & \multicolumn{4}{c}{ISIC Skin Lesion Dataset}\\
\cline{2-9}
   \multicolumn{1}{c}{} 
 & \multicolumn{2}{c}{Modification}
 & \multicolumn{2}{c}{Accuracy}    
 & \multicolumn{2}{c}{Modification}    
 & \multicolumn{2}{c}{Accuracy}\\
\hline
 \phantom{-}Optimization\phantom{-} & $L_2$ & $L_{\infty}$  & IoU & PA & $L_2$ & $L_{\infty}$ & IoU & PA \\
\hline
\multirow{2}{*}{SSM} & $\mathbf{4.60}$ &  $0.22$ & $\mathbf{47\%}$ &  $94\%$ &  $\mathbf{11.76}$ & $0.24$ &  $\mathbf{43\%}$ & $88\%$  \\
~ & $\pm1.76$ &  $\pm0.09$  & $\pm18\%$ &  $\pm2\%$ &  $\pm4.11$ & $0.05$ &  $\pm15\%$ & $\pm2\%$  \\

\cline{1-1}
\multirow{2}{*}{ASM} & $2.82$ &  $0.17$ & $94\%$ &  $99\%$ &  $4.11$ & $0.16$ &  $89\%$ & $98\%$  \\
~ & $\pm1.29$ &  $\pm0.09$  & $\pm7\%$ &  $\pm1\%$ &  $\pm2.23$ & $\pm0.10$ &  $\pm9\%$ & $\pm1\%$  \\

\cline{1-1}
ASM + DPM & $\mathbf{2.47}$ &  $0.17$ & $\mathbf{97\%}$ &  $99\%$ &  $\mathbf{3.88}$ & $0.16$ &  $\mathbf{89\%}$ & $98\%$  \\
(ASMA) & $\pm1.05$ &  $\pm0.09$ & $\pm2\%$ &  $\pm1\%$ &  $\pm1.99$ & $\pm0.09$ &  $\pm10\%$ & $\pm1\%$  \\
\hline
\end{tabularx}
\end{table}

\begin{figure}[t]
\centering
\begin{tikzpicture}
\centering
\node[inner sep=0pt] (whitehead) at (0, 0)  
    {\includegraphics[width=.2\textwidth]{opt_ims/from1.png}};
\node[inner sep=0pt] (whitehead) at (3, 0)  
    {\includegraphics[width=.2\textwidth]{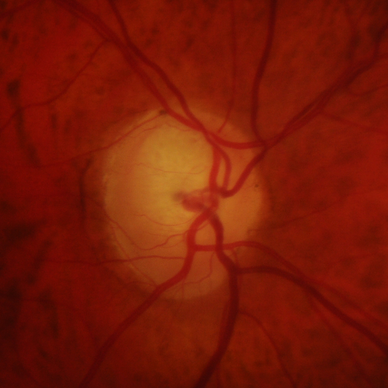}};
\node[inner sep=0pt] (whitehead) at (6, 0) 
    {\includegraphics[width=.2\textwidth]{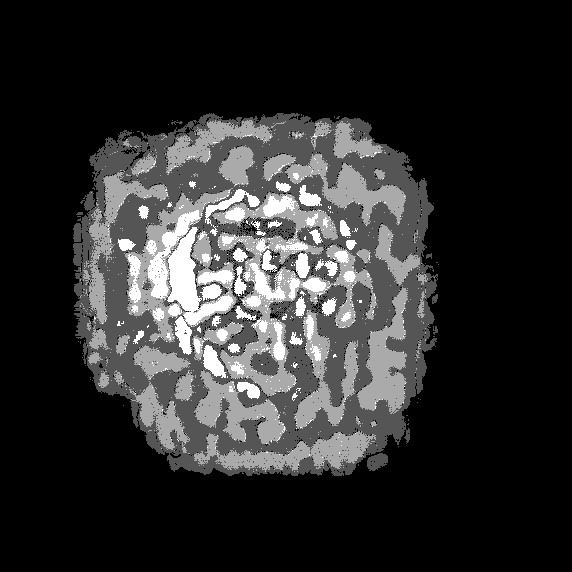}};
\node[inner sep=0pt] (whitehead) at (9, 0)  
    {\includegraphics[width=.2\textwidth]{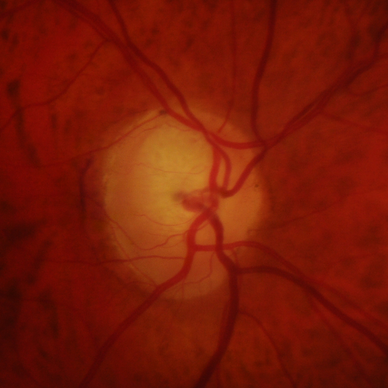}};
\node[inner sep=0pt] (whitehead) at (0, -3)  
    {\includegraphics[width=.2\textwidth]{opt_ims/to.png}};
\node[inner sep=0pt] (whitehead) at (3, -3)  
    {\includegraphics[width=.2\textwidth]{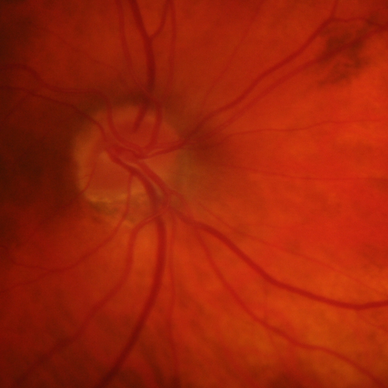}};
\node[inner sep=0pt] (whitehead) at (5.35, -2.35)  
    {\includegraphics[width=.09\textwidth]{opt_ims/sum_im1.png}};
    \node[inner sep=0pt] (whitehead) at (6.65, -2.35)  
    {\includegraphics[width=.09\textwidth]{opt_ims/sum_im2.png}};
    \node[inner sep=0pt] (whitehead) at (5.35, -3.64)  
    {\includegraphics[width=.09\textwidth]{opt_ims/sum_im3.png}};
    \node[inner sep=0pt] (whitehead) at (6.65, -3.64)  
    {\includegraphics[width=.09\textwidth]{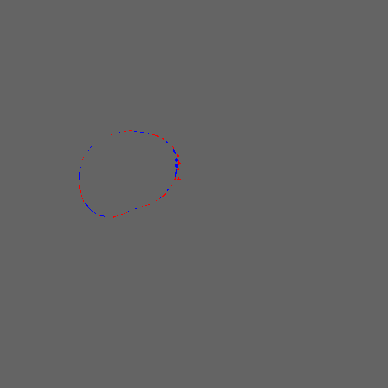}};
\node[inner sep=0pt] (whitehead) at (9, -3) 
    {\includegraphics[width=.2\textwidth]{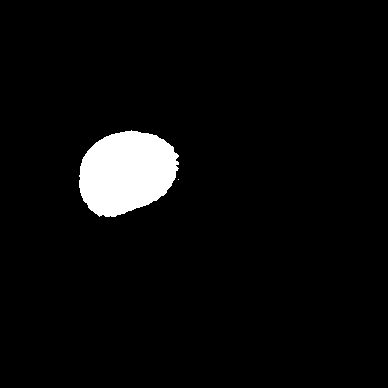}};
    \draw[-{Latex[width=1.5mm]}, line width=0.5mm] (1.75, 0) -- (1.2, 0);
    \draw[-{Latex[width=1.5mm]}, line width=0.5mm] (1.75, -3) -- (1.2, -3);
    \draw[-{Latex[width=1.5mm]}, line width=0.5mm] (6, -1.75) -- (6, -1.2);
    \draw[-{Latex[width=1.5mm]}, line width=0.5mm] (9, -1.25) -- (9, -1.8);
    \node[align=center] at (4.5, 0) {$+$};
    \node[align=center] at (7.5, 0) {$=$};
    \node[align=center] at (0, 1.6) {Segmentation\\mask of (a)};
    \node[align=center] at (3, 1.6) {(a) Source\\\phantom{-}image};
    \node[align=center] at (6, 1.8) {Generated\\perturbation\\(Enhanced $\times$ 100)};
    \node[align=center] at (9, 1.8) {Generated\\adversarial example\\ $L_2 = 2.3 ,\,\, L_{\infty}=0.16$};
    \node[align=center] at (0, -4.9) {Segmentation\\mask of (b)\\(Target adversarial mask)};
    \node[align=center] at (3, -4.7) {(b) Target\\\phantom{-}image};
    \node[align=center] at (6, -4.9) {Adaptive\\optimization\\masks};
    \node[align=center] at (9, -5.1) {Predicted\\segmentation for the\\adversarial example\\$IOU = 98\% ,\,\, PA=99\%$};
\end{tikzpicture}
\caption{An example optimization of an adversarial example for segmentation using ASMA. Best viewed in color.}
\label{fig:adv_example}
\end{figure}

\section{Experiments}
\label{Experiments_Section}
Table~\ref{tab1} presents quantitative results on the viability of the proposed algorithm for adversarial example generation. Specifically, Table~\ref{tab1} shows the degree of perturbation in terms of $L_2$ and $L_{\infty}$ distances, as well as the mask accuracy of the produced adversarial examples in terms of IoU and PA, hereby detailing the influence of the incremental updates (i.e., SSM, ASM, and DPM) that were discussed in Section~\ref{Adaptive Mask Segmentation Attack}. The values that can be found in Table~\ref{tab1} have been determined by calculating the mean and the standard deviation obtained from the optimization of $1000$ adversarial examples. For each of those optimizations, the target adversarial mask is randomly selected among the masks of the other samples, so to have a realistic target in terms of medical image segmentation. To find an optimal perturbation multiplier when DPM is not incorporated, we used $\alpha \in \{1e{-}8, \,1e{-}7, \,1e{-}6, \,1e{-}5\}$; when DPM is incorporated into the optimization, we use $\beta \in  \{1e{-}6, \,5e{-}6, \,1e{-}5\}$ and $\tau = 1e{-}7$. Results are listed for the experiment that achieves the highest average IoU score throughout the $1000$ generated adversarial examples for the selected set of parameters.

As can be observed from Table~\ref{tab1}, the proposed method, when all enhancements introduced in Section~\ref{Adaptive Mask Segmentation Attack} are incorporated, achieves $97\%$ and $89\%$ IoU overlap with the target mask that was used for initiating the optimization, with $L_2$ perturbations as low as $2.47$ and $3.88$ for the glaucoma optic disc and the ISIC skin lesion segmentation problems, respectively. For a more intuitive understanding, an $L_2$ perturbation of $3.88$ corresponds to a modification of less than $1\%$ of the images used in this study.  Our experiments showed that, using ASMA, tuning the perturbation multiplier parameters $\beta$ and $\tau$  is not a difficult task, as we were able to achieve high IOU overlap with low $L_2$ and $L_{\infty}$ perturbations with only $3$ possible combinations of $\beta$.

Apart from the quantitative results presented in Table~\ref{tab1}, we also provide a qualitative overview of the approach used to generate adversarial examples in Figure~\ref{fig:adv_example}. Specifically, Figure~\ref{fig:adv_example} shows the optimization procedure, a generated adversarial example, and the resulting predicted segmentation for a sample taken from the glaucoma optic disc dataset~\cite{Glucoma_seg}. As can be seen, our algorithm is able to completely change the prediction to the desired output mask (taken from another sample in the dataset) with an IoU success rate of $98\%$.

\section{Conclusions and Future Work}
\label{Conclusion and Directions for Future Work}

In this paper, we demonstrated that deep learning-based models for medical image segmentation are vulnerable to attacks using adversarial examples, hereby focusing on skin lesion and glaucoma optic disc segmentation. Specifically, we introduced the Adaptive Mask Segmentation Attack, a novel algorithm that is able to produce adversarial examples with realistic prediction masks that have been altered to be misclassified, using perturbations mostly invisible to the human eye. The source code of our adversarial attack, as well as additional examples, can be found at \url{https://github.com/utkuozbulak/adaptive-segmentation-mask-attack}.

Although we were able to observe similar results on different models, as well as transferability of our generated adversarial examples to other models, we leave it to future work to perform a more detailed analysis and to examine corresponding novel defense mechanisms against attacks that leverage realistic prediction masks.

\section*{Acknowledgements}
The research activities described in this paper were funded by Ghent University Global Campus, Ghent University, imec, Flanders Innovation \& Entrepreneurship (VLAIO), the Fund for Scientific Research-Flanders (FWO-Flanders), and the EU.

\bibliographystyle{splncs04}
\bibliography{2019_02}

\begin{thebibliography}{10}
\providecommand{\url}[1]{\texttt{#1}}
\providecommand{\urlprefix}{URL }
\providecommand{\doi}[1]{https://doi.org/#1}

\bibitem{arnab2018robustness}
Arnab, A., Miksik, O., Torr, P.H.: On the robustness of semantic segmentation
  models to adversarial attacks. In: Proceedings of the IEEE Conference on
  Computer Vision and Pattern Recognition. pp. 888--897 (2018)

\bibitem{CW_Attack}
Carlini, N., Wagner, D.A.: Towards evaluating the robustness of neural
  networks. CoRR  \textbf{abs/1608.04644} (2016)

\bibitem{DBLP:journals/corr/CarliniW17}
Carlini, N., Wagner, D.A.: Adversarial examples are not easily detected:
  Bypassing ten detection methods. CoRR  \textbf{abs/1705.07263} (2017)

\bibitem{Cordts2016Cityscapes}
Cordts, M., Omran, M., Ramos, S., Rehfeld, T., Enzweiler, M., Benenson, R.,
  Franke, U., Roth, S., Schiele, B.: The cityscapes dataset for semantic urban
  scene understanding. In: Proc. of the IEEE Conference on Computer Vision and
  Pattern Recognition (CVPR) (2016)

\bibitem{pascal_voc}
Everingham, M., Eslami, S.M.A., Van~Gool, L., Williams, C.K.I., Winn, J.,
  Zisserman, A.: The pascal visual object classes challenge: A retrospective.
  International Journal of Computer Vision  \textbf{111}(1),  98--136 (Jan
  2015)

\bibitem{medical_classification_adv}
Finlayson, S.G., Kohane, I.S., Beam, A.L.: Adversarial attacks against medical
  deep learning systems. arXiv preprint arXiv:1804.05296  (2018)

\bibitem{ISIC_dataset}
Gutman, D., Codella, N.C.F., Celebi, M.E., Helba, B., Marchetti, M.A., Mishra,
  N.K., Halpern, A.: Skin lesion analysis toward melanoma detection: {A}
  challenge at the international symposium on biomedical imaging {(ISBI)} 2016,
  hosted by the international skin imaging collaboration {(ISIC)}. CoRR
  \textbf{abs/1605.01397} (2016)

\bibitem{medical_seg1}
Heimann, T., Meinzer, H.P.: Statistical shape models for 3d medical image
  segmentation: a review. Medical image analysis  \textbf{13}(4),  543--563
  (2009)

\bibitem{IFGS}
Kurakin, A., Goodfellow, I., Bengio, S.: Adversarial examples in the physical
  world. CoRR  \textbf{abs/1607.02533} (2016)

\bibitem{Glucoma_seg}
Pena-Betancor, C., Gonzalez-Hernandez, M., Fumero-Batista, F., Sigut, J.,
  Medina-Mesa, E., Alayon, S., de~la Rosa, M.G.: Estimation of the relative
  amount of hemoglobin in the cup and neuroretinal rim using stereoscopic color
  fundus images. Investigative ophthalmology \& visual science  \textbf{56}(3),
   1562--1568 (2015)

\bibitem{DBLP:journals/corr/RonnebergerFB15}
Ronneberger, O., Fischer, P., Brox, T.: U-net: Convolutional networks for
  biomedical image segmentation. In: Medical Image Computing and
  Computer-Assisted Intervention -- MICCAI 2015. pp. 234--241. Springer
  International Publishing, Cham (2015)

\bibitem{LBFGS}
Szegedy, C., Zaremba, W., Sutskever, I., Bruna, J., Erhan, D., Goodfellow, I.,
  Fergus, R.: Intriguing properties of neural networks. CoRR
  \textbf{abs/1312.6199} (2013)

\bibitem{xie2017adversarial}
Xie, C., Wang, J., Zhang, Z., Zhou, Y., Xie, L., Yuille, A.: Adversarial
  examples for semantic segmentation and object detection. In: Proceedings of
  the IEEE International Conference on Computer Vision. pp. 1369--1378 (2017)

\end{thebibliography}
\end{document}